\documentclass[prl,twocolumn,showpacs]{revtex4}

\usepackage{times}
\usepackage{bm}
\usepackage{tabularx}
\usepackage[dvips]{graphicx}
\usepackage{amsmath}
\usepackage{bm}
\usepackage{natbib}
\usepackage{color}
\usepackage{epstopdf}
\begin{document}
\title{Spin Transport in Hydrogenated Graphene}
\author{David Soriano$^{1}$, Dinh Van Tuan,$^{1,2}$, Simon M.-M. Dubois$^{3}$, Martin Gmitra$^{4}$, Aron W. Cummings$^{1}$, Denis Kochan$^{4}$, Frank Ortmann,$^{5,6}$, Jean-Christophe Charlier$^{3}$, Jaroslav Fabian$^{4}$ and Stephan  Roche,$^{1,7}$}
\affiliation{$^1$ICN2 - Institut Catal\`a de Nanoci\`encia i Nanotecnologia,
 Campus UAB, 08193 Bellaterra (Barcelona), Spain\\
 $^2$ Department of Physics, Universitat Aut\'{o}noma de Barcelona, Campus UAB, 08193 Bellaterra, Spain\\
$^3$ Universite catholique de Louvain (UCL), Institute of Condensed Matter and Nanosciences (IMCN), Chemin des \'etoiles 8, 1348 Louvain-la-Neuve, Belgium\\
$^4$ Institute for Theoretical Physics, University of Regensburg, 93040 Regensburg, Germany\\
$^5$ Institute for Materials Science and Max Bergmann Center of Biomaterials, Technische Universit\"{a}t Dresden, 01062 Dresden, Germany\\
 $^6$ Dresden Center for Computational Materials Science, Technische Universit\"{a}t Dresden, 01062 Dresden, Germany\\
$^7$ICREA, Instituci\'{o} Catalana de Recerca i Estudis Avan\c{c}ats,
  08070 Barcelona, Spain}
\date{\today}

\begin{abstract}
In this review we discuss the multifaceted problem of spin transport in hydrogenated graphene from a theoretical perspective. The current experimental findings suggest that hydrogenation can either increase or decrease spin lifetimes, which calls for clarification. We first discuss the spin-orbit coupling induced by local $\sigma-\pi$ re-hybridization and ${\bf sp}^{3}$ C-H defect formation together with the formation of a local magnetic moment. First-principles calculations of hydrogenated graphene unravel the strong interplay of spin-orbit and exchange couplings. The concept of magnetic scattering resonances, recently introduced \cite{Kochan2014} is revisited by describing the local magnetism through the self-consistent Hubbard model in the mean field approximation in the dilute limit, while spin relaxation lengths and transport times are computed using an efficient real space order N wavepacket propagation method. Typical spin lifetimes on the order of 1 nanosecond are obtained for 1 ppm of hydrogen impurities (corresponding to transport time about 50 ps), and the scaling of spin lifetimes with impurity density is described by the Elliott-Yafet mechanism. This reinforces the statement that magnetism is the origin of the substantial spin polarization loss in the ultraclean graphene limit. 
\end{abstract} 

\pacs{72.80.Vp, 73.63.-b, 73.22.Pr, 72.15.Lh, 61.48.Gh} 
\maketitle

\textit{Introduction}.-Remarkable electronic and transport properties of graphene (see for instance \cite{Foa2014}) can be further tailored towards higher functionality through the use of chemical functionalization \cite{Loh201}, irradiation (defect formation) \cite{Krasheninnikov2007} or structural patterning (such as creating a nanomesh superlattice) \cite{Yang2014}. Among the wealth of possible modifications, hydrogenation, ozonization or fluorination have demonstrated the tunability of graphene from a weakly disordered semimetal to a wide band-gap or Anderson insulator depending on the nature and density of impurity atoms, varying typically from $0.001\%$ to few percent \cite{Elias2009, Cheng2010, Moser2010}. Together with Raman spectroscopy which provides an estimate of defect density, analysis of the low-temperature conductance behaviour and weak localization regime enable evaluation of the main transport length scales (such as the mean free path and localization length) \cite{Moser2010,Roche2012}. Surprisingly, strongly hydrogenated graphene, with anomalously large Ioffe-Regel ratio of $1/k_{F}\ell_{e}\sim 240$ still exhibits quantum Hall effect features in the high magnetic field regime \cite{Guillemette2013}.
  
Hydrogen defects are particularly interesting since, together with a resonant scattering state created locally in space, the breaking of the sublattice symmetry entails the formation of a zero-energy mode and a local magnetic moment on the order of 1 Bohr magneton for an isolated hydrogen adatom \cite{Yazyev2007}. Meanwhile, the coupling between different induced magnetic moments is either ferromagnetic or antiferromagnetic, depending on whether the H-adatoms correspond to the same or to different sublattices of the graphene lattice, respectively. Theoretical calculations have reported specific magnetoresistance signals for specific long range magnetic ordering situations \cite{Soriano2011}, which could be realized by substrate-induced chemical reactivity as proposed in Ref. \cite{Hemmatiyan2014}.

Graphene spintronics has attracted a lot of attention since the pioneering demonstration that spin could be efficiently injected and propagated over long distances at room temperature  \cite{Tombros2007}. This has opened an opportunity for the development of lateral spintronics \cite{han2011,Guimaraes2012,Maasen2012,Dlubak2012,Zomer2012,Valenzuela2013,Guimaraes2014},  that would benefit from the unique features of graphene, such as a high mobility and remarkable electronic and transport features like relativistic energy dispersion, Klein tunneling phenomenon,... \cite{Han2014}.  In Table 1, we compile the typical values obtained for charge and spin transport in a variety of graphene devices, from graphene supported on silicon oxide or boron nitride to suspended graphene. One observes that the variation of mobility varies by up to two orders of magnitude, whereas spin lifetime seems difficult to relate to the quality of the material.

\noindent

\onecolumngrid

\begin{table}    
\caption{Charge/spin transport parameters in graphene on various substrates \label{specs}}    
\begin{tabularx}{0.6\linewidth}{|X|c|c|c|c|c|}                                                             
\hline                 
\hline 
Substrate                                                       &$\mu$ (cm$^2/Vs$)   &$D_c$ ($m^2s^{-1}$) &$D_s$ ($m^2s^{-1}$)&$\tau_s (ps)$     & $\lambda_{sf}(\mu m)$ \\
\hline
$SiO_2$\cite{Tombros2007}   & $2\times10^3$      & $2\times10^{-2}$   & $2\times10^{-2}$  & $\sim 100$       & $1.5-2$\\
$SiO_2$\cite{han2011}  & $1-3\times10^3$      & $-$                  &$1.3\times10^{-2}$ & $\sim 500$       & $2.4$\\
Suspended\cite{Valenzuela2013}  & $3\times10^5$      & $0.1-0.2$                &$0.05-0.1$& $\sim 150$ 
 & $4.7$\\
$SiC$\cite{Maasen2012} & $2\times10^3$          &  $2\times10^{-2}$           & $4\times10^{-4}$ & $2300$& $0.5-1$\\
$SiC$\cite{Dlubak2012}      & $17\times10^3$        & $-$                & $-$              & $-$      & $>100$\\
$hBN$\cite{Zomer2012}  & $4\times 10^4$ & $0.05$& $0.05$ & $200$& $4.5$\\
{\small $hBN/G/hBN$}\cite{Guimaraes2014}& $2.3\times 10^4$ & $0.03-0.1$& $0.05$ & $3000$& $12$\\
\hline
\hline                                                                                   
\end{tabularx}                                                                                
\end{table}    

\twocolumngrid

\noindent
\vspace{10pt}

Understanding spin transport in hydrogenated graphene is a challenging and important issue. The underlying goal is to use a low enough level of hydrogenation to preserve a sizable transport signal while also inducing local or long range magnetic ordering (ideally a ferromagnetic ground state), and evaluate how spin diffusion is affected by interaction between itinerant spins and local magnetic moments. Ultimately, one could envision spin devices such as spin valves that do not use ferromagnetic materials to inject and detect spins, but rather utilize the varying signal response of “magnetized graphene” to manipulate the spin degree of freedom and engineer logic functions, a long-standing quest of spintronics \cite{Fabian2007}.

From that perspective, the first experimental result showing some interaction between spin diffusion and magnetic moments produced by H adatoms was reported in 2012 by McCreary and coworkers \cite{McCreary2012}.  Low-temperature spin transport measurements on graphene spin valves (T=15K) were shown to exhibit a dip in the non-local spin signal as a function of the external magnetic field. This effect was tunable with hydrogen density, and was related to spin relaxation spin relaxation by exchange coupling with paramagnetic moments. Spin lifetimes were estimated to be $\tau_{s}\sim 400-500$ps in absence of hydrogen, and were slightly enhanced upon increasing the H-adatom density up to $n_{H}\sim 0.1\%$), and accompanied by a reduction of charge mobility by one order of magnitude to $\mu\sim 500{\rm cm^{2}/Vs}$. 

A subsequent experiment also reported a surprising increase of spin lifetime with hydrogen density, with values enhanced by a factor of two after hydrogenation of the samples for  $n_{H}\sim 0.02\%$ \cite{Wojtaszek2013}. The upscaling of $\tau_{s}$ with hydrogen density suggests a Dyakonov-Perel relaxation mechanism \cite{Dyakonov1971}. In these experiments, the effect of magnetic moments is proposed to be strong enough to counteract the expected increase of spin-orbit coupling, which would be expected to produce shorter spin lifetimes with increasing H-adatom density.

In 2013, J. Balakrishnan and coworkers \cite{Balakrishnan2013} reported experimental evidence for a room temperature spin Hall effect (SHE) in weakly hydrogenated graphene, with non-local spin signals up to $100\Omega$ (orders of magnitude larger than in metals). The non-local SHE revealed by Larmor spin-precession measurements was assigned to a “colossal enhancement” of the spin-orbit interaction induced by H-adatoms for density in the range $n_{H}\sim 0.01\%-0.05\%$, mobilities $\mu\sim 14.000-900{\rm cm^{2}/Vs}$) and spin lifetime on the order of $\tau_{s}\sim 100$ps.  The spin-orbit interaction was estimated to be about 2.5 meV, one order of magnitude higher than {\it ab initio} calculations \cite{Gmitra2013}. The observation of the large SHE-signal was exclusively assigned to an enhancement of spin-orbit coupling by H-adatoms, which limits spin lifetimes. It is however not clear whether the previously observed contribution of magnetic moments in spin transport \cite{McCreary2012, Wojtaszek2013} remains marginal or not for the formation and strength of the SHE signal, and which mechanism (Dyakonov-Perel \cite{Dyakonov1971} or Elliott-Yafet \cite{ElliotYafet}) can explain the variety of conflicting experimental data.

In this context, it is of prime importance to clarify the specific impact of H-adatoms on spin relaxation in graphene, and to quantify the relative contribution of spin-orbit coupling and magnetic effects. A first fundamental advance in this direction has been made by Kochan, Gmitra and Fabian \cite{Kochan2014} who introduced a new relaxation mechanism in hydrogenated graphene driven by resonant scattering by magnetic impurities. By neglecting the spin-orbit coupling effect, spin lifetimes of $\tau_{s}\sim 150$ps were estimated for 1 ppm Hydrogen at room temperature and for large contribution of electron-hole puddles. 

\section{Local magnetic moments and spin-orbit coupling from first-principles}
\noindent

Here we remind the main ingredients and predictions of the model introduced in Ref.\cite{Kochan2014}, and present new results obtained with a more generalized microscopic approach to study spin dynamics and relaxation effects induced by magnetic moments. In the latter model, we describe the local moment formation using the self-consistent Hubbard model on hydrogenated  graphene \cite{Soriano2011}. This model is shown to correctly reproduce {\it ab initio} results for spin splittings obtained for small supercells, whereas it provides, by a scaling analysis, a more extended spreading of local magnetic moments around the H impurity when compared with the analytic model used in \cite{Kochan2014}. Next, the Hubbard tight-binding parameters are implemented into a real space wavepacket propagation method which gives direct access to spin dynamics and spin relaxation effects \cite{Dinh2014}. 

\subsection{First-principles calculations of local magnetism and 
spin-orbit coupling effects in hydrogenated graphene}

When a hydrogen atom is chemisorbed, graphene locally undergoes a structural deformation, breaking the $sp^2$ symmetry in favor of an $sp^3$-like hybridization. After {\it ab initio} structural optimization, the hydrogenated carbon atom is found to be slightly displaced out of the plane ($\sim 0.5\AA$), forming a C-H bond of length $1.13$$\AA$. As measured experimentally, the $\sigma - \pi$ rehybridization induced by the hydrogen adatom comes with the formation of a local magnetic moment of the order of 1$\mu_B$. Recently, spin-unpolarized first-principles calculations suggested that the local $sp^3$ distortion and its resulting pseudospin inversion asymmetry induce a giant enhancement of the spin-orbit coupling (SOC) \cite{Gmitra2013}. In this section, spin-polarized first-principles calculations are used to investigate the interplay between the local magnetic moment and the enhanced spin-orbit interaction induced by hydrogen adatoms. Using unconstrained spinors to represent the one-particle wave-functions, the respective contributions to the energy bands splitting of both SOC and electronic exchange are computed. The spin textures associated with the low energy electronic spectrum are also investigated. 

Hydrogenated graphene in the dilute limit is represented by a large supercell (5x5x1) containing a single hydrogen defect, leading to a hydrogen concentration of $\sim 2\%$. The out-of-plane dimension of the cell ensures a distance $> 15$$\AA$ between neighboring graphene planes in order to avoid interaction between periodically repeated images.  The one particle Hamiltonian is computed within the framework of non-collinear spin-polarized density functional theory as implemented in the Vienna Ab initio Simulation Package (VASP) \cite{Kresse1996}. The projector augmented wave method is used to expand the one-particle wave-functions up to an energy cutoff of $600$eV \cite{Blochl1994}. The eigenstates of the self-consistent Hamiltonian are populated according to an electronic temperature of 300K.  Electronic exchange and correlation are treated within the generalized gradient approximation by means of the PBE functional \cite{Perdew1996}. Integrals over the Brillouin zone are performed using a 6x6x1 Monkhorst-Pack grid of k-points. The geometry is fully optimized until remaining atomic forces and stresses are lower than $0.01$eV/$\AA$ and $0.01$eV/$\AA^2$ respectively. Various magnetization axes, either parallel or perpendicular to the graphene plane,  have been considered. The different configurations are energetically equivalent, i.e. the  computed total energies vary by less than 1 $\mu$eV. In what follows, we focus on the most symmetrical and experimentally relevant configuration where the magnetization axis is chosen as the direction perpendicular to the graphene plane. 

\begin{figure}[h!]
\begin{center}
\includegraphics[width=0.92\linewidth]{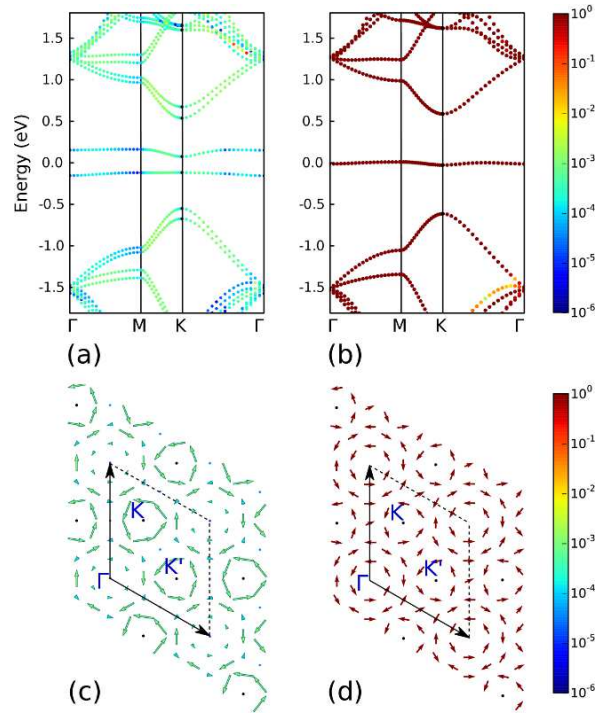}
\caption{(a,b) {\it Ab initio} band structures of hydrogenated graphene in the dilute limit ($\sim 2\%$) including spin-orbit coupling calculated within both (a) spin-polarized and (b) spin-unpolarized frameworks. The color scale is associated with the in-plane component of the spin expectation values. Hence, the blue and red extrema of the color scale correspond to eigenstates whose spin is parallel and perpendicular to the graphene plane. Black dots are determined by symmetry. (c,d) In plane projection of the spin texture of the highest valence band in both the spin-polarized (c) and -unpolarized (d) frameworks. The arrows indicate the in-plane projections of the spin directions (rescaled for eye convenience). The color scale indicates the actual values of the in-plane projections. Red and blue arrows respectively correspond to spin momenta that are parallel and perpendicular to the graphene plane.}
\label{Figure1}
\end{center}
\end{figure}

The computed low-energy band structure is depicted in Fig.\ref{Figure1} (a). The hydrogen impurity opens an energy gap of $1.24$eV and produces two nearly-flat bands around the Fermi energy, accounting for the computed total magnetic moment of $0.99\mu_B$. In this configuration, the spin degeneracy is broken and the energy splitting of the bands originates from both electronic exchange and spin-orbit interaction. For the sake of comparison, the low energy spectrum computed within a spin-unpolarized framework is also reported in Fig.\ref{Figure1} (b). We note that spin-unpolarized means that the density matrix is constrained to be diagonal with same values for both spin-directions. In this case, the energy splitting of the bands only originates from the spin orbit interaction. At the scale of Fig.\ref{Figure1}(b), the splitting is not visible to the eye.

\begin{figure}[h!]
\begin{center}
\includegraphics[width=0.99\linewidth]{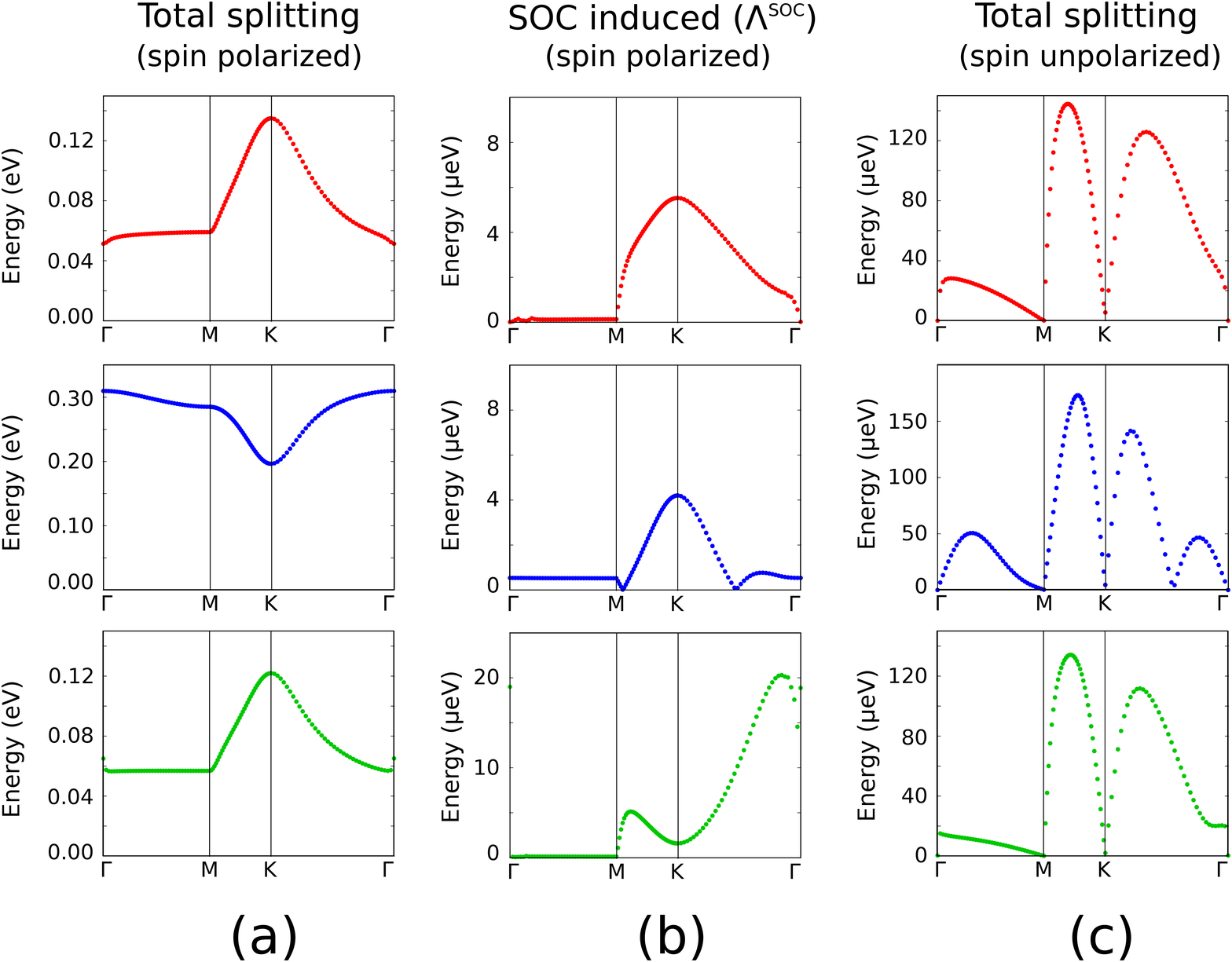}
\caption{Energy splittings of the conduction (red curves), impurity (blue curves) and valence (green curves) bands along high-symmetry lines in reciprocal space. (a) Total energy splitting within the spin -polarized framework. (b) Contribution of SOC to the energy splitting within  the spin polarized framework (i.e. difference between the energy splittings computed respectively with and without accounting for the contribution of the spin-orbit interaction to Hamiltonian). (c) Total energy splitting within the spin-unpolarized framework (i.e. only arising from SOC as the electronic density is constrained to be spin-up polarized).}
\label{Figure2}
\end{center}
\end{figure}

The computed energy splittings corresponding to the low-energy bands are
reported in Fig.\ref{Figure2} (a,c). The energy splitting computed within
the spin unpolarized framework [see Fig.\ref{Figure2} (c)] is in perfect
agreement with previous estimates \cite{Gmitra2013}. In order to clarify
the impact of the spin-orbit interaction, the SOC-induced splitting
($\Lambda^{SOC}$) here defined as the difference between the energy splittings computed with and without the spin-orbit interaction is also reported in Fig.\ref{Figure2} (b). $\Lambda^{SOC}$ is
shown to be at least two orders of magnitude smaller than the energy splitting
associated with the stabilization of the local magnetic moment by the
electronic exchange (Fig.\ref{Figure2} (a)). This highlights the prevailing
role of electronic exchange in ruling the low-energy spectrum in the presence of 
local magnetic moments. The comparison of Fig.\ref{Figure2} (b) and (c) 
further emphasizes the interplay between spin-orbit and exchange couplings.

The reduction of SOC-induced spin splitting due to exchange coupling is explained by considering a simple effective Hamiltonian. Consider electrons with momentum $k$ along $x$ and energy $\varepsilon_k$,  experiencing Rashba SOC $\alpha$ which is in general momentum dependent. If the electrons also feel an exchange coupling $\Delta$
with a magnetic moment oriented along $z$, as is our case, the Hamiltonian is
\begin{equation}
H = \varepsilon_k + \Delta \sigma_z + \alpha(k) \sigma_y. 
\end{equation}
The energy splitting of the eigenvalues of this operator is 
\begin{equation}
2\sqrt{\Delta^2 + \alpha^2}.
\end{equation}
In the absence of exchange, the splitting is linear in $\alpha$, since Rashba SOC splits two degenerate levels. In the presence of exchange, and for realistic $\Delta \gg \alpha$, as in our case, the Rashba contribution to the splitting is only second order in $\alpha$ ($\frac{\alpha^2}{\Delta}$). This rough estimate gives for SOC of 100 $\mu$eV and $\Delta$ of 0.1 eV, a Rashba coupling contribution to the splitting of approximately 0.1 $\mu$eV, significantly below what is observed in Fig.\ref{Figure2} (b). It is then likely that the difference in the calculated exchange coupling with and without SOC is due to the Rashba dependence of the exchange itself. This could be linear with $\alpha$, with a numerical factor. The understanding of this dependence in terms of simple phenomenological models is still elusive and is currently the object of further investigation.

The spin texture of the valence band is illustrated in Fig.\ref{Figure1} (c). When the electronic density matrix is constrained to be spin-unpolarized, the frustration of the electronic exchange "artificially" leads to a highly non-collinear spin texture entirely ruled by SOC (see Fig.\ref{Figure1} (d)). On the contrary, in the unconstrained framework, the stabilization of the magnetic moment dominates the energetics of the low energy spectrum and the spin texture is found to be nearly collinear with the spin-orbit interaction only accounting for $\sim0.1\%$ of the deviation from collinearity (Fig.\ref{Figure1} (c)). While the local rehybridization of graphene induces an enhancement of the spin-orbit interaction,  the magnetic properties of graphene as well as its spin texture are quasi-exclusively ruled by the electronic exchange in the presence of hydrogen-induced local magnetic moments. Hence, the contribution of the spin-orbit interaction to the Hamiltonian is neglected in the following discussion on the spin lifetimes in hydrogenated graphene. 

\section{One-orbital mean-field Hubbard approximation}
\begin{figure}[h!]
\begin{center}
\includegraphics[width=0.75\linewidth]{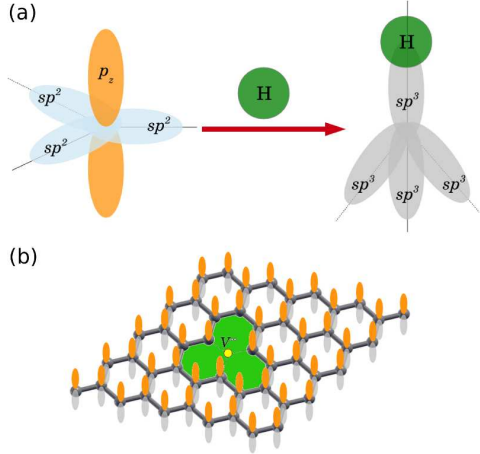}
\caption{(a) Schematic representation of the $sp^2$ hybridization 
breakdown upon hydrogen adsorption, equivalent to the creation of a single vacancy. 
(b) Single-vacancy 5x5 supercell used in the mean-field Hubbard calculations with periodic boundary conditions. The $p_z$ orbital at the center of the vacancy (yellow site inside the green region) is removed by adding a large on-site potential ($V^\infty$) to simulate hydrogen adsorption.}
\label{Figure3}
\end{center}
\end{figure}

\noindent
The electronic structure of graphene is modeled by a nearest-neighbor tight-binding model with a single $p_z$-orbital per site. When a H atom is adsorbed on top of a carbon (C) atom, the $sp^2$-symmetry is locally broken, and the electron from the C $p_z$ orbital is removed from the $\pi$ bands to form a $\sigma$ bond with the H atom (Fig.\ref{Figure3} (a)). Figure~\ref{Figure3}(b) shows a 5x5 supercell where the absence of a $p_z$-orbital in the center of the green region (yellow site) represents hydrogen adsorption. To remove the $p_z$ orbital, we use a sufficiently large on-site potential $V^\infty \approx 10^4$ eV. To properly describe magnetism in hydrogenated graphene, we introduce on-site Coulomb repulsion between electrons with opposite spins by means of the Hubbard model in its mean-field approximation 
\begin{equation}
\label{MFHM}
\mathcal{H} = -\gamma_0\sum_\mathrm{\langle i,j \rangle,\sigma} c_\mathrm{i,\sigma}^\dagger c_\mathrm{j,\sigma} + 
U\sum_\mathrm{i} \left (n_\mathrm{i,\uparrow} \langle n_\mathrm{i,\downarrow} \rangle +
 n_\mathrm{i,\downarrow} \langle n_\mathrm{i,\uparrow} \rangle \right ),
\end{equation} 
%
%
\noindent where $t$ is the first-neighbors hopping term, $c_\mathrm{i,\sigma}^\dagger$ ($c_\mathrm{j,\sigma}$) is the creation (annihilation) operator in the lattice site $i$ ($j$) with spin $\sigma$,  $U$ is the on-site Coulomb repulsion, and $\langle n_\mathrm{i,\downarrow}\rangle$, $\langle n_\mathrm{i,\uparrow} \rangle$ are the converged expectation values of the occupation numbers for spin-down and spin-up electrons, respectively. The ratio $U/\gamma_0 = 1$ has been chosen to accurately reproduce first-principles calculations.

Hydrogen defects in graphene may induce sublattice symmetry breaking leading to the appearance of zero-energy modes in the density of states (DOS), as predicted by Inui {\it et al.}~\cite{Inui1994}, which are mainly localized around the impurities~\cite{Ugeda,CastroNeto}. These zero-energy states become spin-polarized upon switching on the Coulomb repulsion, leading to semi-local $S=1/2$ magnetic moments with a staggered spin density primarily localized on one sublattice~\cite{Yazyev2007,Palacios2008,Soriano2010}. For a finite concentration of defects, the long range ordering of magnetic moments is dictated by the type of sublattice functionalization, being co-polarized or ferromagnetic (FM) for the same sublattice and counter-polarized or antiferromagnetic (AFM) otherwise. The total spin $S$ of the macroscopic ground state is given by the excess of magnetic moments on one chosen sublattice~\cite{Lieb1989}, although $S=0$ is the most likely value on simple statistical grounds (equal H occupation of both sublattices). Here we consider the dilute limit so that long range magnetic states are neglected and we get local paramagnetic (PM) impurities.

\begin{figure}[h!]
\begin{center}
\includegraphics[width=\linewidth]{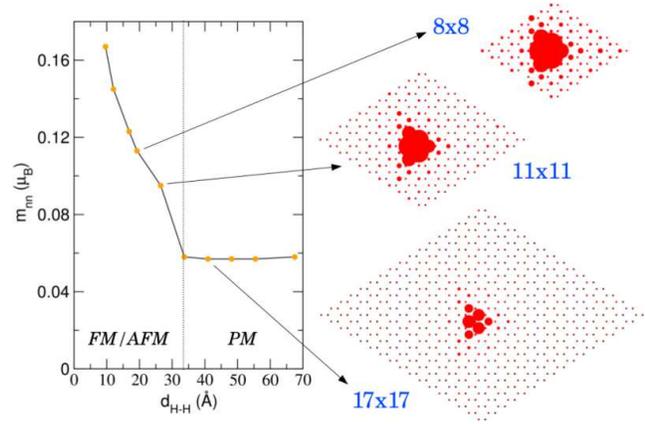}
\caption{Evolution of magnetic moments on first neighbor atom for increasing distance between impurities ($d_{H-H}$). The value converges apparently for $d_{H-H} \geq 33.74$ \AA ~where long range magnetic interactions start to vanish precluding the formation of FM or AFM ordering and leading to isolated paramagnetic defects. On the right side of the figure, we show the magnetic moments $m_i$ corresponding to each lattice site for different supercell sizes.}  
\label{Figure4}
\end{center}
\end{figure}

In order to reach a pure paramagnetic state, we have studied the evolution of the local magnetic moments $m_i = 1/2\left(n_{i,\uparrow} - n_{i,\downarrow}\right)$ with the varying supercell size (Fig.\ref{Figure4}). In particular, Fig.\ref{Figure4} (left panel) shows the values of the magnetic moment only at nearest neighbor sites of the defect $m_{nn}$. As the supercell size is increased and the defects move away from each other, $m_{nn}$ decreases quickly and converge for a supercell size 14x14, which corresponds to $0.25$ \% of adsorbed H and $d_{H-H} = 33.74$ \AA. This decay is related to the overlap between wavefunctions corresponding to neighboring defects. For strongly overlapping states (large concentration of H), long-range interactions induce a FM state between magnetic moments located in neighboring cells which increases the values of $m_i$. When the overlap is very small (diluted limit), long-range interactions have no influence on the local magnetism leading to almost constant $m_i$ values around the defect (PM state).       

\begin{figure}[h!]
\begin{center}
\includegraphics[width=\linewidth]{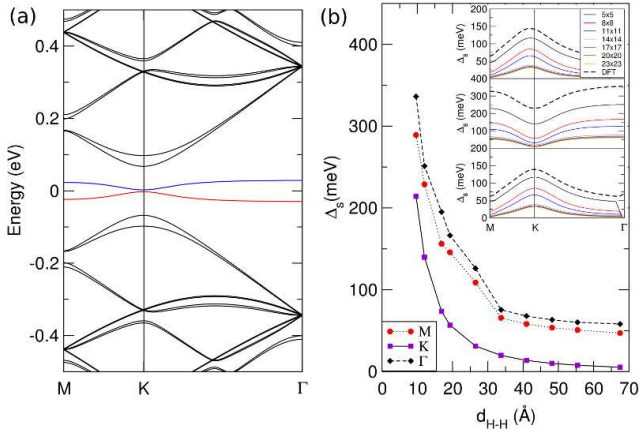}
\caption{(a) Band structure corresponding to $0.06 \%$ of H adsorbed (28x28 supercell). The spin degeneracy is broken at the Fermi energy ($E=0$) due to the formation of local magnetic fields around impurities. (b) Evolution of $\Delta_s$ at three symmetry points K, $\Gamma$ and M, for increasing values of the distance between impurities ($d_{H-H}$). The inset shows the evolution of $\Delta_s$ along the M-K-$\Gamma$ path in the hexagonal Brillouin zone of the valence (bottom), impurity (center) and conduction (top) bands. Dashed lines correspond to DFT results.}
\label{Figure5}
\end{center}
\end{figure}

Figure \ref{Figure5}(a) shows the band structure of a single hydrogenated 28x28 graphene supercell, corresponding to $0.06 \%$ of adsorbed H, calculated using the mean-field Hubbard model (Eq. \ref{MFHM}). The splitting ($\Delta_s$) of the mid-gap state formed during H adsorption is plotted in Fig. \ref{Figure5}(b) for different supercell sizes at different $k$-points. At the K point  $\Delta_s$ decays as $\propto d_{H-H}^{-1.9}$, while at the $\Gamma$ and M-points $\Delta_s \propto d_{H-H}^{-1}$. These results confirm the existence of very small splittings at tiny concentrations of H (of the order of 1ppm). In order to compare the values of the energy splitting of spin polarized bands obtained using the mean-field Hubbard model with published DFT results~\cite{Kochan2014}, we have plotted $\Delta_s$ along M-K-$\Gamma$ path in the inset of Fig. \ref{Figure5}(b). Bottom, middle and top panels correspond to the valence, mid-gap and conduction bands respectively. Although DFT results for a 5x5 supercell (dashed lines) show a slightly higher splitting of the spin polarized bands with respect to our results, this methodology nicely capture the main physics of magnetic resonances induced by H adatoms in graphene. 
 
\section{Spin dynamics and relaxation}

We now investigate the spin dynamics and relaxation phenomena in hydrogenated graphene by comparing two complementary theoretical approaches and contrasting our results with state-of-the-art experimental data. 

\subsection{Single impurity limit}

First, we consider the magnetic scattering problem in the single impurity limit, which can be simplified by considering only the spin-flip processes stemming from the exchange coupling on the resonant scatterer (H atom) site. This was done in details in Ref.~\cite{Kochan2014} on the basis of the Hamiltonian $\hat{\mathcal{H}}=\hat{\mathcal{H}}_0+\hat{\mathcal{H}}_{S}$ which involves the usual $p_{z}$-orbital Hamiltonian for the pristine graphene $\hat{\mathcal{H}}_0=-\gamma_{0}\sum_{\langle ij\rangle }c_i^+c_j$ together with the term $\hat{\mathcal{H}}_{S}$ describing graphene-hydrogen chemisorption including the interaction of electron spin $\hat{\bf s}_h$ and impurity moment $\hat{\bf S}$ (exchange coupling)
\begin{equation}
\begin{aligned}
\hat{\mathcal{H}}_{S}&=\sum\limits_{\sigma}\epsilon_h\, h_\sigma^\dagger h_\sigma +
T\left[ h_\sigma^+c^{\phantom{+}}_{C_H,\sigma}+c_{C_H,\sigma}^+h_\sigma\right]\\
&-J\,\hat{\mathbf{s}}_h\cdot\hat{\mathbf{S}}
\end{aligned}
\label{RGl}
\end{equation}
Here $h_\sigma^+ $ ($h^{\phantom{+}}_\sigma $) and $c_{m,\sigma}^+$ ($c^{\phantom{+}}_{m,\sigma}$) are fermionic operators creating (annihilating) electron with spin $\sigma$ on the hydrogen and graphene carbon site $m$, respectively. The orbital chemisorption parameters entering the Hamiltonian $\hat{\mathcal{H}}_{S}$ are extracted from the {\it ab-initio} calculation in Refs.~\cite{Kochan2014,Kochan2014b}---the hydrogen on-site energy $\epsilon_h=0.16$~eV and graphene-hydrogen hybridization $T=7.5$~eV.
The exchange term $-J\,\hat{\mathbf{s}}_h\cdot\hat{\mathbf{S}}$ describes an effective interaction of the local magnetic moment $\hat{\mathbf{S}}$ induced on the hydrogen site with an itinerant electron spin $\hat{\mathbf{s}}_h$ when this hops into the hydrogen $|h\rangle\equiv h^+|0\rangle$-level that hybridizes with the graphene host.
Both, $\hat{\mathbf{S}}$ and $\hat{\mathbf{s}}_h$ are vectors of spin $\tfrac{1}{2}$-operators (in our definition without conventional $\tfrac{\hbar}{2}$ factor) and $J$ is a constant with the dimension of energy. For the exchange we take $J=-0.4\,\mathrm{eV}$ since this value is consistent with a more detailed
parametrization of the magnetic impurity model as discussed in Ref.~\cite{Kochan2014}. In fact the precise value and the sign of $J$ are not important as long as $|J|$ is greater that the orbital resonance width (for the case of hydrogen the resonance width is $4$~meV).
In the independent electron-impurity picture (we do not discuss Kondo physics), the exchange term $-J\,\hat{\mathbf{s}}_h\cdot\hat{\mathbf{S}}$ can be diagonalized introducing the singlet ($\ell=0$) and triplet ($\ell=1$) composite spin states $|\ell,m_\ell\rangle$ ($m_\ell=-\ell,\dots,\ell$). Transforming the initial Hamiltonian $\hat{\mathcal{H}}_{S}$, Eq.~(\ref{RGl}), to the new spin basis we arrive at
\begin{equation}
\begin{aligned}
\hat{\mathcal{H}}_{S}&=
h^+h\otimes\sum\limits_{\ell,m_\ell}\bigl[\epsilon_h-J(4\ell-3)\bigr]|\ell,m_\ell\rangle\langle \ell,m_\ell|\\
&+T\left[ h^+c^{\phantom{+}}_{C_H}+c_{C_H}^+h\right]\otimes\sum\limits_{\ell,m_\ell}|\ell,m_\ell\rangle\langle \ell,m_\ell|\,.
\end{aligned}
\label{RG2}
\end{equation}
Down-folding the hydrogen $|h\rangle$-state by means of the L\"{o}wdin transformation we get for each spin component $\ell, m_\ell$ an independent effective delta-function problem
\begin{equation}
\begin{aligned}
\hat{\mathcal{H}}_{S}^{\mathrm{eff}}(E)&=\sum\limits_{\ell,m_\ell}\,\mathrm{V}_\ell(E)\,c_{C_H}^+c^{\phantom{+}}_{C_H}\otimes|\ell,m_\ell\rangle\langle \ell,m_\ell|\,,
\label{RG3}
\end{aligned}
\end{equation}
located on the hydrogenated carbon site $C_H$ with the energy dependent coupling
\begin{equation}
\mathrm{V}_\ell(E)=\frac{T^2}{E-\epsilon_h+J(4\ell-3)}\,.
\label{RG4}
\end{equation}
From here it is straightforward to compute T-matrix elements for Bloch states $|\boldsymbol{\kappa}\rangle$ and $|\boldsymbol{\kappa'}\rangle$ of the unperturbed graphene (to shorthand the notation $\boldsymbol{\kappa}$ comprises band index and crystal momentum). Assuming the Bloch states are normalized to graphene unit cell the result is as follows
\begin{equation}
\mathrm{T}_{\boldsymbol{\kappa},\ell, m_\ell|\boldsymbol{\kappa}',\ell', m_{\ell'}}(E)=\delta_{\ell\ell'}\delta_{m_\ell m_{\ell'}}
\frac{\mathrm{V}_\ell(E)}{1-\mathrm{V}_\ell(E)G_{0}(E)}\,,
\label{RG5}
\end{equation}
where $\mathrm{V}_\ell(E)$ is given by Eq.~(\ref{RG4}) and $G_0(E)$ is Green's function per site and spin for the unperturbed graphene,~i.e.,
\begin{equation}
G_0(E)\simeq\frac{E}{W^2}\ln{\Bigl|\frac{E^2}{W^2-E^2}\Bigr|}-i\pi\,\frac{|E|}{W^2}\Theta(W-|E|)\,.
\label{RG6}
\end{equation}
The above Green's function is valid near the graphene charge neutrality point in the energy window from -1\,eV to 1\,eV, where the linearized bandwidth $W=\sqrt{\sqrt{3}\pi}\gamma_{0}\simeq 6$~eV. For practical reasons we need relaxation rates $1/\tau_{\sigma\sigma'}(E)$ that characterize spin-conserving ($\sigma=\sigma'$) and spin-flipping ($\sigma\neq\sigma'$) processes in graphene at the given Fermi energy $E$ in presence of magnetic active impurities. For that we take into account all scattering processes $|\boldsymbol{\kappa},\sigma,\Sigma\rangle\rightarrow|\boldsymbol{\kappa'},\sigma',\Sigma'\rangle$ at the given energy $E\equiv E(\boldsymbol{\kappa})$ with the requested incoming and outgoing electron (hole) spins $\sigma$ and $\sigma'$ and allowed impurity spins $\Sigma$ and $\Sigma'$ (a charge carrier in graphene flips its spin only if the magnetic moment does the same to conserve the total angular momentum).
To get $1/\tau_{\sigma\sigma'}(E)$ we start with the transition rate $W_{\boldsymbol{\kappa}\sigma\Sigma|\boldsymbol{\kappa'}\sigma'\Sigma'}$ and trace out the $\Sigma$-spin degrees of freedom corresponding to magnetic moment $\hat{\mathbf{S}}$ induced on the hydrogen site, then we integrate over all outgoing momenta $|\boldsymbol{\kappa'}\rangle$ and finally we average the result over all incoming states $|\boldsymbol{\kappa}\rangle$ at the given Fermi energy $E$.
This can be done because the angular dependence of the T-matrix is trivial, it does not depend on the angle of $\boldsymbol{\kappa}$ neither of $\boldsymbol{\kappa'}$.
Assuming the distribution of magnetic moments is dilute and they are in average spin-unpolarized (e.g.~by interaction with phonons) the result is as follows
\begin{equation}
\frac{1}{\tau_{\sigma\sigma'}(E)}=\eta \tfrac{2\pi}{\hbar}\nu_0(E) f_{\sigma\sigma'}\Bigl[\tfrac{\mathrm{V}_1(E)}{1-\mathrm{V}_1(E)G_0(E)},\tfrac{\mathrm{V}_0(E)}{1-\mathrm{V}_0(E)G_0(E)}\Bigr]\,.
\label{RG7}
\end{equation}
Here $\eta$ is the concentration of hydrogen impurities per carbon atom, $\nu_0(E)=-\tfrac{1}{\pi}\,\mathrm{Im}\,G_0(E)$ is the graphene density of states per atom and spin, $V_\ell(E)$ for $\ell=0,1$ is given by Eq.~(\ref{RG4}) and function $f_{\sigma\sigma'}[x,y]$ is defined by
\begin{equation}\label{RG8}
f_{\sigma\sigma'}[x,y]=
\frac{1}{2}\delta_{\sigma\sigma'}\bigl|x\bigr|^2+
\frac{1}{8}\bigl|x+(\sigma\cdot\sigma')\,y\bigr|^2\,.
\end{equation}
The function $f_{\sigma\sigma'}$ originates from the decomposition of the actual spin states $|\sigma,\Sigma\rangle$ and $|\sigma',\Sigma'\rangle$ with respect to the singlet-triplet basis $|\ell,m_\ell\rangle$ after the tracing out the spin states of the induced magnetic moment. The symbol $\sigma\cdot\sigma'$ entering its definition equals 1 (-1) for the parallel (antiparallel) spin alignments and $\bigl|\phantom{x}\bigr|$ stands for the absolute value.

Knowing the partial rates $1/\tau_{\sigma\sigma'}(E)$ we can define the \emph{spin-relaxation rate} $1/\tau_s(E)=1/\tau_{\downarrow\uparrow}(E)+1/\tau_{\uparrow\downarrow}(E)$ and \emph{momentum-relaxation rate} $1/\tau(E)=1/\tau_{\uparrow\uparrow}(E)+1/\tau_{\uparrow\downarrow}(E)$ at zero temperature as functions of Fermi energy.
From Eq.~(\ref{RG7}) we immediately see that both $1/\tau_s$ and $1/\tau$ should go to zero at the charge neutrality point since there are
no states that can participate in scattering, $\nu_0(0)=0$. Secondly, the denominators $1-\mathrm{V}_0(E)G_0(E)$ and $1-\mathrm{V}_1(E)G_0(E)$ that enter the function $f_{\sigma\sigma'}$ can minimize at certain energies $E_0$ and $E_1$ and this would manifests as two sharp peaks (singlet and triplet one) in the spin and momentum relaxation rates. Both mentioned features are clearly seen in Fig.~\ref{Figure7}.

To account for finite temperature effects one should thermally broaden $1/\tau_s$ implementing rate equations for the graphene charge-carriers that obey Fermi-Dirac statistics. The concise formula for the spin-relaxation rate at finite temperature $T$ then becomes
\begin{equation}\label{RG9}
\frac{1}{\tau_s(E,T)} = \frac{\sum\limits_{k}\bigl[- \partial_{\epsilon_k} \mathrm{F}(\epsilon_k,E, T)\bigr]
1/\tau_s(\epsilon_k)}{\sum\limits_{k} \bigl[- \partial_{\epsilon_k} \mathrm{F}(\epsilon_k,E, T)\bigr]}\,,
\end{equation}
where $\mathrm{F}(\epsilon_k,E, T)$ is the equilibrium Fermi-Dirac distribution at temperature $T$ and Fermi energy $E$. Experiments show that charged impurities and electron-hole puddles within the sample still further affect system's Fermi energy. To account for all that effects one usually convolves $1/\tau_s(E,T)$ by the Gaussian kernel with standard deviation $\sigma_b$,
\begin{equation}\label{RG10}
\frac{1}{\tau_s(E,T)} = \frac{1}{2\sqrt{\pi}\sigma_b}\int\limits_{-\infty}^{\infty}\mathrm{d}\epsilon\,\frac{1}{\tau_s(\epsilon,T)}\,
\exp{\left(-\frac{(E-\epsilon)^2}{2\sigma_b^2}\right)}\,.
\end{equation}
Depending on the sample quality the typical value of $\sigma_b$ to be taken for best fits is about 70-110\,meV (see discussion next section). 

\subsection{Tight-binding Model of hydrogenated graphene}

In our second approach to spin relaxation, we study the spin dynamics using the following tight-binding Hamiltonian,

\begin{eqnarray}
\hat{\mathcal{H}}_{S_{r}} &=&\epsilon_h\sum_{m}h_m^+h_m +T
\sum_{\langle mi\rangle }h_m^+c_i \\
&+& \sum_{i}J_ic_i^+\sigma_zc_i\nonumber
\label{Hamil2}
\end{eqnarray}

which includes the self-consistent Hubbard terms describing isolated magnetic moments (dilute limit), and a random distribution of H-adatoms in the ppm range. The long range nature of magnetic moment induced by hydrogen is involved in this Hamiltonian by considering up to  ninth-nearest-neighbor exchange coupling term $J_i$. Spin-orbit coupling is found to yield negligible corrections to the results, so it is neglected in the following. The spin dynamics are investigated by computing the time-dependence of the spin polarization defined by \cite{Dinh2014}

\begin{equation}
{\bf S}(E,t)=\frac{{\rm Tr}\left[\delta(E-\hat{H})\boldsymbol{\hat{\mathbf s}}(t)\right]}{{\rm Tr}\left[\delta(E-\hat{H})\right]}\nonumber
\end{equation}

\noindent
where $\boldsymbol{\hat{\mathbf s}}(t)=e^{\frac{i\hat{ H} t}{\hbar}} \boldsymbol{\hat{\mathbf s}} e^{\frac{-i\hat{ H} t}{\hbar}}$. Using random phase states to perform the trace efficiently, we get
\begin{equation}
{\bf S}(E,t)=\frac{1}{2\Omega\rho(E)}\langle\psi(t)\lvert\delta(E-\hat{H})\boldsymbol{\hat{\mathbf s}}+\boldsymbol{\hat{\mathbf s}}\delta(E-\hat{H})\rvert\psi(t)\rangle\nonumber
\end{equation}
\noindent
and the initial wavepacket can be prepared in an arbitrary spin polarization as
 \begin{equation}
\vert \Psi_{RP} \rangle=\frac{1}{\sqrt{N}} \sum_{i=1}^N 
 \begin{pmatrix}
\cos \left(\frac{\theta_i}{2}\right)\\ e^{i\varPhi_i}\sin \left(\frac{\theta_i}{2}\right)
\end{pmatrix}
  e^{2i\pi\alpha_i} | i  \rangle \nonumber
  \end{equation}

In what follows, the wavepackets are prepared (at $t=0$) with an in-plane spin polarization and their time-dependent propagation and $S_{x}(E,t)$ are evaluated applying the evolution operator (Schr\"odinger equation) to the wavepackets . In absence of spin-orbit interaction, the spin dynamics is influenced by the existence of local magnetism, with H-related magnetic moments pointing out-of-plane. 

\begin{figure}[t!]
\begin{center}
\includegraphics[width=1\linewidth]{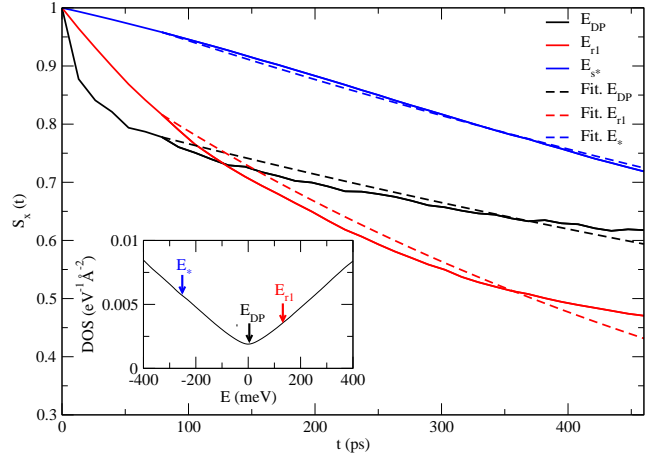}
\caption{(Main frame) time-dependent polarization for initial in-plane spin polarized wavepackets at selected energies (see inset). Numerical fits using an exponential damping are shown (dashed lines) on top of numerical simulations start from elapsed propagation time $t>75$ps. Inset: total density of state for the case of 1ppm of hydrogen on graphene.}
\label{Figure6}
\end{center}
\end{figure}

Fig.\ref{Figure6} shows the time-evolution of the spin polarization at three different energies, namely the Dirac point, an energy close to the expected resonance and some high energy value (see inset). The curves first exhibit an sudden drop (especially for the Dirac point) followed by an exponential decay, which dictates the values of the spin relaxation times $\tau_s$ using $S_x(t)=S_x(t_0)e^{-t/\tau_s}$ (fitting the numerical results from $t_0=75$ps). The initial fast decay of $S_{x}(t)$ is understood as follows.  We study the propagation of wavepackets which (for computational efficiency) are at time $t=0$ in a random phase state, that is extended through the whole system. For smooth disorder, this method well captures the main transport length scales as described in Ref.\cite{Foa2014}.  Here however, the hydrogen adatoms (or vacancies) produce strong magnetic moments localized around the impurity. This introduces some transient decay of the initial spin polarization which is not representative of the long time exponential behavior. 

\subsection{Spin relaxation times in hydrogenated graphene}

\begin{figure}[h!]
\begin{center}
\includegraphics[width=1\linewidth]{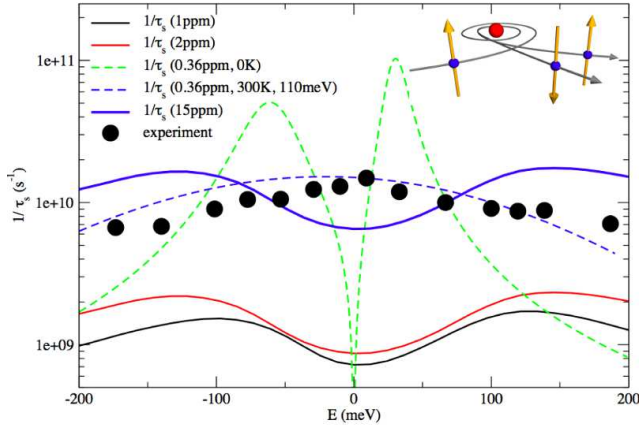}
\caption{(Main frame) Energy-dependence of $\tau_{s}$ derived from the T-matrix of the single impurity model at zero temperature (green dashed lines); and at 300 K broadened by puddles with energy fluctuations of 110 meV (blue dashed lines). Same quantity obtained from the real space spin propagation method for 1 ppm (solid black line), 2 ppm (solid red line) and 15 ppm (solid blue line) of H adatoms. Inset: schematic of the magnetic resonance process at the origin of enhanced spin relaxation.}
\label{Figure7}
\end{center}
\end{figure}

In Fig.\ref{Figure7}, the inverse spin relaxation times obtained for the two models are superimposed.  Within the single impurity approximation, two sharp {\it magnetic resonances} are clearly formed  (for singlet and triplet states), at which $1/\tau_{s}$ is maximum at zero temperature (green dashed lines). This approximation of $\tau_{s}$ however diverges close to the Dirac point, which disagrees with most low-temperature experimental data. Room temperature spin-relaxation  experimental data can however be reproduced using thermal and charge-puddles broadenings. A choice of 0.36 ppm of H-impurities together with a finite temperature broadening (300 K) and charge density fluctuations of 110 meV are necessary (blue dashed line) to reproduce experimental data (black filled circles) obtained for high-quality graphene spin-valve devices. \cite{Wojtaszek2013}. 

The inverse spin lifetimes obtained by the real space spin propagation method are shown in Fig.\ref{Figure7} for a H-density of 1 ppm (black solid line), 2 ppm (red  solid line) and 15 ppm (blue solid line).  The presence of the two resonances is confirmed by this calculation. At the resonances, the inverse spin lifetime is greater than in the single impurity approximation, which is likely due to the different natures of the adatom models. In the single-impurity adatom model the the magnetic moment is strongly localized and the resonance is sharp, while in the Hubbard model we have an effective defect model with more delocalized magnetism is real space. This broadening reduces the {\it ``residency time"} of the electrons on the {\it magnetic virtual bound state} (as pictured in the inset of Fig.\ref{Figure7}-inset), and thus yields longer spin lifetimes. Indeed for 0.36 ppm of H-impurities, the two models differ to three orders of magnitude in spin lifetime. By further contrasting the real-space spin results to the experimental data  (black dots in Fig.\ref{Figure7}), we extrapolate that a density of 10-20 ppm of H-adatoms can reproduce the range of experimental values of the graphene samples measured in Ref.\cite{Wojtaszek2013}.  

\begin{figure}[t!]
\begin{center}
\includegraphics[width=1\linewidth]{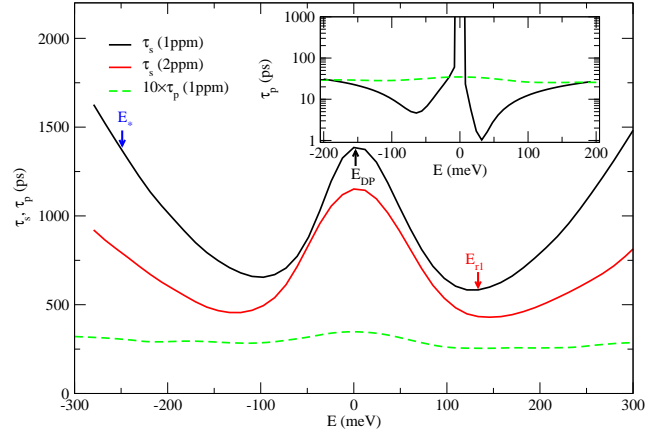}
\caption{Main frame: Energy-dependence of spin (solid lines) and rescaled momentum (dashed line) relaxation times obtained from the real space wavepacket propagation method. Inset: momentum relaxation time derived the from single impurity model \cite{Kochan2014} (solid line) and from the diffusion coefficient behavior (dashed green line).}
\label{Figure8}
\end{center}
\end{figure}

To discuss the relaxation mechanism, we compute the momentum relaxation time $\tau_p$ first from the sum of spin conserving rate and spin-flip rate, which both can be calculated by transforming the singlet and triplet T-matrix amplitudes via composite spin states of electron and impurity \cite{Kochan2014}. The second approach uses the full Hamiltonian $\hat{\mathcal{H}}_{S}$ and the time dependence of the diffusion coefficient $D(E,t)$ \cite{Roche2012}.  For the hydrogen coverage as low as $0.02\%$, we still can observe the saturation of the diffusion coefficient $D(E,t) \longrightarrow D_{max}(E)$ , allowing us to extract $\tau_p$ as
\begin{equation}
\tau_p(E)=\frac{D_{max}(E)}{4v^2(E)}
\end{equation}
where $v(E)$ is the pristine graphene velocity. We use the Fermi golden rule and the expected scaling of $\tau_p(E)\sim 1/n_{H}$, to extrapolate the values for each ppm concentration (see Fig.\ref{Figure8}).  The result is given in Fig.\ref{Figure8} (green solid lines), and shows a weak energy dependence, with slight increase close to the Dirac point, with values in the range of 60-70 ps for 1 ppm of hydrogen impurities. In Fig.\ref{Figure8}(inset), $\tau_p$ for the two models are reported showing some discrepancy close to the Dirac point, as expected from the approximations made. The values match well at high enough energy.

We observe that $\tau_{s}\sim 10\tau_p$, while the scaling of $\tau_{s}$ with impurity density clearly manifests an Elliott-Yafet spin relaxation mechanism, as predicted for such type of impurities \cite{Ochoa2012}. In some experiments, $\tau_{s}$ is found to increase with hydrogen concentration, suggesting differently a Dyakonov-Perel relaxation mechanism \cite{Wojtaszek2013}, whereas others report on the opposite trend associated to the Elliott-Yafet mechanism \cite{Balakrishnan2013}. The origin of such inconsistency remains obscure at that point but could stem (in particular) from the segregation of H-adatoms and the alteration of fundamental effects induced by magnetic resonances.

In conclusion, we investigated the impact of hydrogen adatoms on charge and spin transport in transport in graphene in the dilute limit (down to the ppm limit), for which no long range magnetic ordering develops.  The importance of magnetic resonances as a new spin relaxation mechanism, as pioneered in Ref.\cite{Kochan2014} has been consolidated and further quantified using extended models of more delocalized magnetism, described within the Hubbard Hamiltonian in the mean field approximation. Using efficient real space order N wavepacket propagation methods, spin relaxation times in the nanosecond regime were obtained for 1 ppm hydrogen impurities, and for momentum scattering time few orders of magnitude shorter. More work is needed to describe larger hydrogen densities for which magnetic interaction and long range magnetic ground states could develop and interfere with relaxation effects.

\section{Acknowledgements}
The research leading to these results has received funding from the European Union Seventh Framework Programme under grant agreement number 604391 Graphene Flagship. This work was also funded by Spanish Ministry of Economy and Competitiveness under contracts MAT2012-33911, the Severo Ochoa Program (MINECO SEV-2013-0295). S. R. Thanks the Secretaria de Universidades e Investigaci\'on del Departamento de Econom\'{\i}a y Conocimiento de la Generalidad de Catalu\~{n}a. J.C.C. acknowledges the National Fund for Scientific Research [F.R.S.-FNRS] of Belgium for financial support and the Research Concerted Action on Graphene Nanoelectromechanics (No. 11/16-037). Computational resources have been provided by the supercomputing facilities of the "Consortium des Equipements  de Calcul Intensif" en F\'ed\'eration Wallonie-Bruxelles (CECI), the UCL-CISM, and PRACE.  J.F. acknowledges support from DFG SFB 689 and GRK 1570.

\end{document}